\documentclass[aps,pra,superscriptaddress]{revtex4}
\usepackage{mathrsfs}
\usepackage{amsfonts}
\usepackage{graphicx}
\usepackage{amssymb}
\usepackage{amsmath}
\usepackage[retainorgcmds]{IEEEtrantools}
\begin{document}

\def\cl{\centerline}
\def\bd{\begin{description}}
\def\be{\begin{enumerate}}
\def\ben{\begin{equation}}
\def\benn{\begin{equation*}}
\def\een{\end{equation}}
\def\eenn{\end{equation*}}
\def\benr{\begin{eqnarray}}
\def\eenr{\end{eqnarray}}
\def\benrr{\begin{eqnarray*}}
\def\eenrr{\end{eqnarray*}}
\def\ed{\end{description}}
\def\ee{\end{enumerate}}
\def\al{\alpha}
\def\b{\beta}
\def\bR{\bar\R}
\def\bc{\begin{center}}
\def\ec{\end{center}}
\def\dg{\dagger}
\def\d{\dot}
\def\D{\Delta}
\def\del{\delta}
\def\ep{\epsilon}
\def\g{\gamma}
\def\G{\Gamma}
\def\h{\hat}
\def\iny{\infty}
\def\La{\Longrightarrow}
\def\la{\lambda}
\def\m{\mu}
\def\n{\nu}
\def\noi{\noindent}
\def\Om{\Omega}
\def\om{\omega}
\def\p{\psi}
\def\pr{\prime}
\def\r{\ref}
\def\R{{\bf R}}
\def\ra{\rightarrow}
\def\up{\uparrow}
\def\dn{\downarrow}
\def\lr{\leftrightarrow}
\def\s{\sum_{i=1}^n}
\def\si{\sigma}
\def\Si{\Sigma}
\def\t{\tau}
\def\th{\theta}
\def\Th{\Theta}

\def\vep{\varepsilon}
\def\vp{\varphi}
\def\pa{\partial}
\def\un{\underline}
\def\ov{\overline}
\def\fr{\frac}
\def\sq{\sqrt}
\def\ot{\otimes}
\def\tf{\textbf}
\def\WW{\begin{stack}{\circle \\ W}\end{stack}}
\def\ww{\begin{stack}{\circle \\ w}\end{stack}}
\def\st{\stackrel}
\def\Ra{\Rightarrow}
\def\R{{\mathbb R}}
\def\mf{\mathbf }
\def\bi{\begin{itemize}}
\def\ei{\end{itemize}}
\def\i{\item}
\def\bt{\begin{tabular}}
\def\et{\end{tabular}}
\def\lf{\leftarrow}
\def\nn{\nonumber}
\def\va{\vartheta}
\def\wh{\widehat}
\def\vs{\vspace}
\def\Lam{\Lambda}
\def\sm{\setminus}
\def\ba{\begin{array}}
\def\ea{\end{array}}
\def\bd{\begin{description}}
\def\ed{\end{description}}
\def\lan{\langle}
\def\ran{\rangle}
\def\l{\label}
\def\mb{\mathbb}
\def\ti{\times}
\def\mc{\mathcal}
\def\v{\vec}
\large

\preprint{}
\title{ Geometric measure of quantum discord and total quantum correlations in a N-partite quantum state.}

\date{\today}
\author{Ali Saif M. Hassan}
\email{alisaif73@gmail.com}
 \affiliation{Department of Physics, University of Amran, Amran, Yemen}

\author{Pramod S. Joag}
\email{pramod@physics.unipune.ac.in}
\affiliation{Department of Physics, University of Pune, Pune, India-411007.}

\date{\today}
\begin{abstract}

Quantum discord, as introduced by Olliver and Zurek [Phys. Rev. Lett. \textbf{88}, 017901 (2001)], is a measure of
the discrepancy between quantum versions of two classically equivalent expressions for mutual information and is found to be useful in quantification and application of quantum correlations in mixed states. It is viewed as a key resource present in certain quantum communication tasks and quantum computational models without containing much entanglement. An early step toward the quantification of quantum discord in a quantum state was by
Dakic, Vedral, and Brukner [Phys. Rev. Lett. 105,190502 (2010)] who introduced a geometric measure of quantum discord and
derived an explicit formula for any two-qubit state. Recently, Luo and Fu [Phys. Rev. A \textbf{82}, 034302 (2010)] introduced a generic form of the geometric measure of quantum discord for a bipartite quantum state. We extend these results and find generic forms of the geometric measure of quantum discord and total quantum correlations in a general N-partite quantum state. Further, we obtain computable exact formulas for the geometric measure of quantum discord and total quantum correlations in a N-qubit quantum state. The exact formulas for the $N$-qubit quantum state are experimentally implementable.

\noi PACS numbers: 03.65.Ud;75.10.Pq;05.30.-d
 \end{abstract}

\maketitle

\textbf{I. INTRODUCTION}\\

In quantum information theory, the problem of characterization of correlations present in a quantum state
has been a fundamental problem generating intense research effort during the last two decades \cite{hor,guh}. Correlations in quantum states, with far reaching  implications for quantum information processing, are usually studied in the entanglement-versus-separability framework \cite{wer,hor}. However, some results showed that quantum correlations cannot be only limited to entanglement, because separable quantum states can also have correlations which are responsible for the improvements of some quantum tasks that cannot be achieved by classical means [4-10]. An alternative classification for correlations based
on quantum measurements has arisen in recent years and also plays an important role in quantum information theory [11-14]. This is the quantum-versus-classical paradigm for correlations. The first attempts in this direction were made by Ollivier and Zurek \cite{olli} and by Henderson and Vedral \cite{hend}, who studied quantum correlations from a measurement perspective and introduced quantum discord as a measure of quantum correlations which has generated increasing interest [17-47]. Recently, it was suggested that the quantum discord  $D(\rho)$ can be expressed alternatively as the minimal loss of correlations caused by the non-selective von Neumann projective measurement given by the set of orthogonal $1D$ projectors $\{\Pi_i^a \}$ acting on one part of the system \cite {fu},

\ben  \l{e1}
D(\rho) = \min_{\Pi^a} \{I(\rho)-I(\Pi^a(\rho))\},
\een
where
\[\Pi^a(\rho) = \sum_{i} (\Pi_i^a\otimes I^b)\rho (\Pi_i^a\otimes I^b).\]
Here the minimum is over von Neumann measurements $\Pi^{a} = \{\Pi_i^a \}$ on a part say $a$ of a bipartite system $ab$ in a state $\rho$ with reduced density operators $\rho^{a}$ and $\rho^{b}$ and $\Pi^a(\rho)$ is the resulting state after the measurement. $I(\rho) = S(\rho^{a})+S(\rho^{b})-S(\rho)$ is the quantum mutual information, $S(\rho)=-tr(\rho\ln \rho)$ is the von Neumann entropy and $I^{b}$ is the identity operator on part $b.$   

The definition of quantum discord in terms of quantum mutual information has the disadvantage that it is very difficult to generalize to the multipartite case \cite{pm}. We can overcome this hurdle by introducing a geometric measure of quantum discord as the distance of the given state to the closest classical quantum  (or the zero discord) state (see Eq.(\r{e6})). Dakic et al. \cite{dakic} introduced a geometric measure of quantum discord given by  
\ben  \l{e2}
D(\rho)=\min_{\chi \in \Omega_0}||\rho-\chi||^2,
\een
where $\Omega_0$ denotes the set of zero-discord states and $||\rho-\chi||^2 := tr(\rho-\chi)^2$
is the square norm in the Hilbert-Schmidt space of linear operators acting on the state space of the system.

Dakic et al \cite{dakic} also obtained an easily computable exact expression for the geometric measure of quantum discord for a two qubit system, which can be described as follows. Consider a two-qubit state $\rho$ expressed in its Bloch representation (see section III) as
\begin{IEEEeqnarray}{rCl}
\rho & = & \frac{1}{4}\big{(} I^a\otimes I^b+\sum_{\alpha=1}^3(x_{\alpha} \sigma_{\alpha} \otimes I^b+I^a \otimes y_{\alpha} \sigma_{\alpha} ) \nn \\
  && +\sum_{\alpha,\beta=1}^3 t_{\alpha\beta} \sigma_{\alpha} \otimes \sigma_{\beta} \big{)},\nn
\end{IEEEeqnarray}
$\{\sigma_{\alpha}\}$ being the Pauli operators. Then its geometric measure of quantum discord is given by \cite{dakic}
\ben   \l{e3}
D(\rho) = \frac{1}{4} (||x||^2 + ||T ||^2 - \lambda_{max}).
\een
Here $\vec{x} := (x_1,x_2,x_3)^t$ and $\vec{y} := (y_1,y_2,y_3)^t $ are coherent (column) vectors for single qubit reduced density operators,
$T = (t_{\alpha\beta} )$ is the correlation matrix, and $\lambda_{max}$ is the largest eigenvalue of the
matrix $\vec{x}\vec{x}^t + T T^t .$ The norms of vectors and matrices are the Euclidean norms, for example, $||x||^2 :=\sum_{\alpha} x^2_{\alpha} .$ Here and throughout this article, the superscript $t$ denotes transpose of vectors and matrices and by the norm of any tensor we mean its Euclidean norm, that is, the square of the norm of a tensor is the sum of squares of its elements. \\

In this paper
\bi

\item We obtain a generic form for the quantum discord $D_k$ (corresponding to the Von Neumann measurement on the $k$th part) in a $N$-partite quantum state (Section II, Theorem 1).

\item We give a formula for quantum discord $D_k$ in a $N$-qubit quantum state, which is exactly computable as well as experimentally implementable (Section III, Theorem 2).

\item We apply the exact formula for $D_k$ (obtained in section III) to some multiqubit states (Section IV).   

\item We give a generic form for the total quantum correlations in a bipartite state which can be exactly computed and experimentally implemented for a two-qubit state (Section V).

\item We give a generic form for the total quantum correlations in a $N$-partite state which can be exactly computed and experimentally implemented for a $N$-qubit state (Section VI).    \\

\ei

Finally, we summarize in section VII.\\         

\textbf{II. QUANTUM DISCORD IN A $N$-PARTITE STATE}\\ 
 
 Consider a multipartite system $\mathcal{H}=\mathcal{H}^{1}\otimes \mathcal{H}^{2} \otimes \cdots \otimes \mathcal{H}^{N}$ with $dim(\mathcal{H}^{m}) = d_m, \; m=1,2,\cdots,N$. Let $L(\mathcal{H}^{m})$ be the Hilbert-Schmidt space of linear operators on $\mathcal{H}^{m}$ with the Hilbert-Schmidt inner product
$$\langle X^{(m)}|Y^{(m)}\rangle := tr X^{(m)\dag} Y^{(m)}.$$

We can define  The Hilbert-Schmidt space $L(\mathcal{H}^{1}\otimes \mathcal{H}^{2} \otimes \cdots \otimes \mathcal{H}^{N})$ similarly.
Let $\{X^{(m)}_i : i = 1,2, . . . ,d_m^2,\; m=1,2,\ldots,N\}$ be a set of Hermitian operators which constitute orthonormal
bases for $L(H^{m})$, then
$$tr X^{(m)}_i X^{(m)}_j = \delta_{ij},$$
and $\{X^{(1)}_{i_1}\otimes X^{(2)}_{i_2} \otimes \cdots \otimes X^{(N)}_{i_N}\}$ constitutes an orthonormal basis for $L(\mathcal{H}^{1}\otimes \mathcal{H}^{2} \otimes \cdots \otimes \mathcal{H}^{N})$. In particular, any $N$-partite state $\rho_{12\cdots N} \in L(\mathcal{H}^{1}\otimes \mathcal{H}^{2} \otimes \cdots \otimes \mathcal{H}^{N})$ can be expanded as
\ben   \l{e4}
\rho_{12\cdots N}= \sum_{i_1 i_2 \cdots i_N} c_{i_1 i_2 \cdots i_N} X^{(1)}_{i_1}\otimes X^{(2)}_{i_2} \otimes \cdots \otimes X^{(N)}_{i_N}\;;i_m  =1,\ldots,d_m^2\;;m=1,\ldots,N,
\een
with $\mathcal{C}=[c_{i_1 i_2 \cdots i_N}] = [tr(\rho_{12\cdots N} X^{(1)}_{i_1}\otimes X^{(2)}_{i_2} \otimes \cdots \otimes X^{(N)}_{i_N})]$ is a $N$-way array (tensor of order $N$) with size $d_1^2d_2^2\cdots d_N^2 .$ 

We can define the geometric measure of quantum discord for a $N$-partite quantum state corresponding to the von Neumann measurement on the $k$th part as
\ben   \l{e5}
D_{k}(\rho_{12\cdots N})=\min_{\chi_{k}}||\rho_{12\cdots N}-\chi_{k}||^2 ,
\een
where the minimum is over the set of zero discord states $\chi_{k}$ [i.e. $D_{k}(\chi_{k})=0$] \cite{okrasa}. A state $\chi_{k} \in L(\mathcal{H}^{1}\otimes \mathcal{H}^{2} \otimes \cdots \otimes \mathcal{H}^{N})$ is of zero discord if and only if it is a classical-quantum state \cite{lin}
\ben  \l{e6}
\chi_{k}=\sum_{l=1}^{d_k} p_l |l\rangle \langle l|\otimes \rho_{[k]|l} ,
\een
where $[k]$ stands for $1 2 \cdots k-1 k+1 \cdots N$, $\{p_l\}$ is a probability distribution over the terms in the sum, $\{|l\rangle\}$ is an arbitrary
orthonormal basis in $\mathcal{H}^{k}$, and $\{\rho_{[k]|l}\}$ is a set of arbitrary states
(density operators acting on $\mathcal{H}^{1}\otimes \mathcal{H}^{2} \otimes \cdots \mathcal{H}^{k-1} \otimes\mathcal{H}^{k+1}\otimes \cdots \mathcal{H}^{N}$). It follows that the quantum discord corresponding to measurement on different subsystems is different, that is, $D_k(\rho)\ne D_l(\rho)\;;\;k\ne l .$

We need to define a product of a tensor with a matrix, the n-mode product \cite{kold06,lmv00}.
The \emph{n-mode} (\emph{matrix}) \emph{product} of a tensor $\mathcal{Y}$ (of order $N$ and with dimension $J_1\times J_2\times \cdots  J_N$) with a matrix $A$ with dimension $I\times J_n$ is denoted by $\mathcal{Y}\times_n A$. The result is a tensor of size $J_1\times J_2\times \cdots J_{n-1}\times I\times J_{n+1}\times\cdots J_N$ and is defined elementwise by
\ben \l{e7}
(\mathcal{Y}\times_n A)_{j_1 j_2\cdots j_{n-1}i j_{n+1} \cdots j_N}=\sum_{j_n=1}^{J_n} y_{j_1j_2\cdots j_N} a_{ij_n}.
\een

Recently, for a bipartite system $ab$ $(N=2)$ with states in $\mc{H}^a \ot \mc{H}^b ,\;dim(\mc{H}^a)=d_a,\;dim(\mc{H}^b)=d_b ,$ S. Luo and S. Fu introduced the following form of geometric measure of quantum discord \cite{fu}
\ben  \l{e8}
D_a (\rho) = tr(CC^t )- \max_{A} tr(ACC^tA^t ),
\een
where $C = [c_{ij} ]$ is an $d_a^2\ti d_b^2$ matrix and the maximum is taken over all $d_a\times d_a^2$-dimensional isometric matrices $A = [a_{li} ]$ such that $a_{li}  = tr(|l\rangle\langle l|X_i)=\lan l|X_i|l\ran,\;\; l = 1,2,\ldots,d_a\;;i = 1,2,\ldots,d_a^2$ and $\{|l\rangle\}$ is any orthonormal basis in $\mc{H}^a$. we generalize this result to $N$-partite quantum states.

\emph{Theorem 1}. Let $\rho_{12\cdots N}$ be a $N$-partite state defined by Eq.(\r{e4}), then
\ben  \l{e9}
D_k(\rho_{12\cdots N})=||\mathcal{C}||^2-\max_{A^{(k)}}||\mathcal{C}\times_{k}A^{(k)}||^2,
\een
where $\mathcal{C}=[c_{i_1 i_2 \cdots i_N}]$ is defined via Eq.(\r{e4}), the maximum is taken over all $d_{k}\times d_{k}^2$-dimensional isometric matrices $A^{(k)}=[a_{li_k}],$ $A^{(k)}(A^{(k)})^t = I_{k},$ such that $a_{li_k}=tr(|l\rangle\langle l| X^{(k)}_{i_k}), \;l=1,2,\ldots,d_{k};\;i_k=1,2,\ldots,d_{k}^2$ and $\{|l\rangle\}$ is any orthonormal basis for $\mathcal{H}^{k}.$

\emph{Proof}: 

We expand the operator $|l\rangle \langle l|$ occurring in the expression for the zero discord state $\chi_k$ (Eq.(\r{e6})) in the orthonormal basis  $\{X_{i_{k}}^{(k)}\}$ in $L(\mathcal{H}^{k})$ as
\ben  \l{eq801}
|l\rangle \langle l|=\sum_{i_{k}=1}^{d_k^2} a_{li_{k}} X_{i_{k}}^{(k)}, \; l=1,2,\cdots,d_{k};
\een
with
\ben \l{e10}
 a_{li_{k}}=tr(|l\rangle \langle l| X_{i_{k}}^{(k)})=\langle l|X_{i_{k}}^{(k)}|l\rangle,
\een
$\{|l\ran \}$ being any orthonormal basis in $\mc{H}^k .$ Clearly, $\sum_{l=1}^{d_{k}}a_{li_{k}} = tr X_{i_{k}}^{(k)}.$ Arranging the coefficients in a row vector as
$$ \vec{a}_l=(a_{l1},a_{l2},\cdots,a_{ld_{k}^2}),$$
we get, by the Parseval theorem of abstract Fourier transform, 
\ben \l{eq901}
||\vec{a}_l||^2=|||l\rangle \langle l|||^2=1.
\een
Where $||\vec{a}_l||^2= \sum_{i_{k}}a_{li_{k}}^2$. Moreover, the orthonormality of $\{|l\rangle\}$
implies that $\{\vec{a}_l\}$ is an orthonormal set of vectors, and therefore
$A^{(k)} = [a_{li_{k}}]$ is an isometry in the sense that $A^{(k)}(A^{(k)})^t = I_{k}.$

Similarly, because $\{X^{(1)}_{i_1}\otimes X^{(2)}_{i_2}\otimes \cdots X_{i_{k-1}}^{(k-1)} \otimes X_{i_{k+1}}^{(k+1)}\otimes \cdots X_{i_N}^{(N)}\}$ constitutes an orthonormal basis for  $\mathcal{H}^{1}\otimes \mathcal{H}^{2} \otimes \cdots \mathcal{H}^{k-1} \otimes\mathcal{H}^{k+1}\otimes \cdots \mathcal{H}^{N}$, which we call the `$X$ basis', we can expand the operator $p_l\rho_{[k]|l}$ occurring in the expression for the zero discord state $\chi_k$ as
$$p_l\rho_{[k]|l}=\sum_{i_1i_2\cdots i_{k-1} i_{k+1} \cdots i_N} b_{i_1i_2\cdots i_{k-1} l i_{k+1} \cdots i_N} X^{(1)}_{i_1}\otimes X^{(2)}_{i_2}\otimes \cdots \ot X_{i_{k-1}}^{(k-1)} \otimes X_{i_{k+1}}^{(k+1)}\otimes \cdots \ot X_{i_N}^{(N)},$$
$$ l=1,2,\cdots,d_{k};\; i_{m}=1,2,\cdots d_{m}^2;\;m=1,\ldots,k-1,k+1,\ldots,N$$
with $b_{i_1i_2\cdots i_{k-1} l i_{k+1} \cdots i_N}= tr(p_l\rho_{[k]|l} X^{(1)}_{i_1}\otimes X^{(2)}_{i_2}\otimes \cdots \ot X_{i_{k-1}}^{(k-1)} \otimes X_{i_{k+1}}^{(k+1)}\otimes \cdots \ot X_{i_N}^{(N)}).$ 
Then we have, using orthonormality of the $X$ basis,  
\ben \l{eq178}
\sum_{i_1i_2\cdots i_{k-1} l i_{k+1} \cdots i_N} b_{i_1i_2\cdots i_{k-1} l i_{k+1} \cdots i_N}^2=p_l^2 tr \rho^2_{[k]|l} .
\een
In view of Eqs.(\r{e5}) and (\r{e6}), the square norm distance between
$\rho_{12\cdots N}$ and $\chi_{k}$ can be evaluated (using the orthonormality of the bases involved and Eq.(\r{eq178})) as

\benr   \nn
 ||\rho_{12\cdots N}-\chi_{k}||^2 &=&  tr\rho^2_{12\cdots N}-2 tr(\rho_{12\cdots N}\chi_{k})+tr\chi_{k}^2    \nn     \\
  &=& \sum_{i_1i_2\cdots  i_N} c^2_{i_1i_2\cdots  i_N} -2 \sum_{i_1i_2\cdots  i_N} c_{i_1i_2\cdots  i_N}   \nn   \\ 
  &&\sum_l^{d_{k}} p_l \langle
  l|X^{(k)}_{i_{k}}|l\rangle tr(\rho_{[k]|l} X^{(1)}_{i_1}\otimes X^{(2)}_{i_2}\otimes \cdots X_{i_{k-1}}^{(k-1)} \otimes
  X_{i_{k+1}}^{(k+1)}\otimes \cdots \otimes X_{i_N}^{(N)}) +\sum_l^{d_{k}} p_l^2 tr \rho^2_{[k]|l}   \nn    \\ 
  &=& ||\mathcal{C}||^2 - 2  \sum_{i_1i_2\cdots  i_N} c_{i_1i_2\cdots  i_N} \sum_l^{d_{k}} a_{li_{k}} b_{i_1i_2\cdots i_{k-1} l i_{k+1} \cdots i_N}       +\sum_{i_1i_2\cdots i_{k-1} l i_{k+1} \cdots i_N} b_{i_1i_2\cdots i_{k-1} l i_{k+1} \cdots i_N}^2    \nn     \\
  &=&||\mathcal{C}||^2 - \sum_{i_1i_2\cdots i_{k-1} l i_{k+1} \cdots i_N}\big{(} \sum_{i_{k}} c_{i_1i_2\cdots  i_N}a_{li_{k}} \big{)}^2   \nn   \\
  &&+\sum_{i_1i_2\cdots i_{k-1} l i_{k+1} \cdots i_N}\big{(}b_{i_1i_2\cdots i_{k-1} l i_{k+1} \cdots i_N}-\sum_{i_{k}} c_{i_1i_2\cdots  i_N}a_{li_{k}}\big{)}^2 .    \nn     \\
\eenr
  By choosing $b_{i_1i_2\cdots i_{k-1} l i_{k+1} \cdots i_N}=\sum_{i_{k}} c_{i_1i_2\cdots  i_N}a_{li_{k}}$ \cite{com},  the above equation reduces to $$||\rho_{12\cdots N}-\chi_{k}||^2 = ||\mc{C}||^2 - ||\mathcal{C}\times_{k}A^{(k)}||^2 .$$ Since the tensor $\mc{C}$ is determined by the state $\rho_{12\cdots N}$ via Eq.(\r{e4}), we have, using Eq.(\r{e2}),  
\ben
D_{k}(\rho_{12\cdots N}) = \min_{\chi_{k}}||\rho_{12\cdots N}-\chi_{k}||^2 = ||\mathcal{C}||^2- \max_{A^{(k)}}||\mathcal{C}\times_{k} A^{(k)}||^2 , \nn
\een
where the maximum is taken over $A^{(k)}$ specified in the theorem, thus completing the proof.

For a bipartite system, $\mc{C}$ is a $d_1^2\ti d_2^2$ matrix while $A^{(1)}$ and $A^{(2)}$ are $d_1 \ti d_1^2$ and $d_2 \ti d_2^2$ matrices respectively. Using the definition of the n-mode product (eq.(\r{e7})) and the norm of a tensor it follows that 
\ben \l{eq2}
D_1(\rho) = tr(CC^t )- \max_{A^{(1)}} tr(A^{(1)}CC^tA^{(1)t}), \\
\een
and
\ben \l{eq3}
D_2(\rho) = tr(CC^t )- \max_{A^{(2)}} tr(A^{(2)}C^tCA^{(2)t}). \\
\een

Following its definition in Eq.(\r{e1}), it seems more natural and simple to define the geometric measure of quantum discord as
\ben \l{e11}
\overline{D}_{k}(\rho_{12\cdots N})= \min_{\Pi^{k}} ||\rho_{12\cdots N}-\Pi^{k}(\rho_{12\cdots N})||^2 ,
\een
where the minimum is over von Neumann measurements $\Pi^{k}= \{\Pi^{k}_l\} $ on system $\mathcal{H}^{k}$, and $\Pi^{k}(\rho_{12\cdots N}) = \sum_l (I_1\otimes I_2 \otimes \cdots \otimes \Pi^{k}_l \otimes \cdots \otimes I_N) \rho_{12\cdots N} (I_1\otimes I_2 \otimes \cdots \otimes \Pi^{k}_l \otimes \cdots \otimes I_N).$

It is easy to prove that $D_{k}(\rho_{12\cdots N})=\overline{D}_{k}(\rho_{12\cdots N})$, similar to theorem 2 in ref. \cite{fu}. \\

\textbf{III. EXACT FORMULA FOR A $N$-QUBIT STATE}\\

In this section we specialize to the $N$-qubit systems with states in $\mb{C}^2\ot\mb{C}^2\cdots \ot \mb{C}^2$ ($N$ factors). We need the structure of the Bloch representation of density operators, which can be briefly described as follows. Bloch representation of a density operator acting on the Hilbert space of a $d$-level quantum system $\mathbb{C}^d$ is given by
\ben  \l{e12}
 \rho = \frac{1}{d} (I_d + \sum_{\alpha} s_{\alpha} {\tilde{\lambda}_{\alpha}}),
 \een
where the components of the coherent vector $\vec{s},$ defined via Eq.(\r{e12}), are given by $s_{\alpha}=\frac{d}{2}tr(\rho {\tilde{\lambda}}_{\alpha}).$
 Eq.(\r{e12}) is the expansion of $\rho$ in the Hilbert-Schmidt basis $\{I_d,\tilde{\lambda}_{\alpha}; \alpha=1,2,\dots,d^2-1\}$ where $\tilde{\lambda}_{\alpha}$ are the traceless hermitian generators of $SU(d)$ satisfying $tr(\tilde{\lambda}_{\alpha} \tilde{\lambda}_{\beta})=2\delta_{\alpha\beta}$  \cite{mahl}.

In order to give the Bloch representation of a density operator acting on the Hilbert
  space $\mathbb{C}^{2} \otimes \mathbb{C}^{2} \otimes \cdots \otimes \mathbb{C}^{2}$
  of a $N$-qubit quantum system, we introduce the following notation. We use $k_i \; (i=1,2,\cdots)$ to denote a qubit chosen from $N$ qubits, so that  $k_i \; (i=1,2,\cdots)$ take values in the set  $\mathcal{N}=\{1,2,\cdots,N\}$. Thus each $k_i$ is a variable taking values in $\mc{N}.$ The variables $\alpha_{k_i}=1,2,3$ for a given $k_i$ span the set of generators of $SU(2)$ group (except identity) for the $k_i$th qubit, namely the set of Pauli operators $\{\sigma_{1},\sigma_{2},\sigma_{3}\}$ for the $k_i$th qubit. For two qubits $k_1$ and $k_2$ we define

$$\sigma^{(k_1)}_{\alpha_{k_1}} =  (I_2\otimes I_2\otimes \dots \otimes \sigma_{\alpha_{k_1}}\otimes I_2\otimes \dots \otimes I_2)$$
$$\sigma^{(k_2)}_{\alpha_{k_2}}  = (I_2\otimes I_2\otimes \dots \otimes \sigma_{\alpha_{k_2}}\otimes I_2\otimes \dots \otimes  I_2)$$
 \ben \l{e13}
   \sigma^{(k_1)}_{\alpha_{k_1}} \sigma^{(k_2)}_{\alpha_{k_2}}  = (I_2\otimes I_2\otimes \dots \otimes \sigma_{\alpha_{k_1}}\otimes
    I_2\otimes \dots \otimes \sigma_{\alpha_{k_2}}\otimes I_2\otimes I_2) ,
 \een
where  $\sigma_{\alpha_{k_1}}$ and $\sigma_{\alpha_{k_2}}$ occur at the $k_1$th and $k_2$th places (corresponding to $k_1$th and $k_2$th qubits respectively) in the tensor product and are the $\alpha_{k_1}$th and  $\alpha_{k_2}$th generators of $SU(2) ,\alpha_{k_1}=1,2,3\; \mbox{and} \; \alpha_{k_2}=1,2,3$ respectively. Then we can write, for a $N$-qubit state $\rho_{12\cdots N} ,$ 
\begin{multline} \l{e14}
\rho_{12\cdots N}=\fr{1}{2^N} \{\otimes_m^N I_{d_m}+ \sum_{k_1 \in \mathcal{N}}\sum_{\alpha_{k_1}}s_{\alpha_{k_1}}\sigma^{(k_1)}_{\alpha_{k_1}} +\sum_{\{k_1,k_2\}}\sum_{\alpha_{k_1}\alpha_{k_2}}t_{\alpha_{k_1}\alpha_{k_2}}\sigma^{(k_1)}_{\alpha_{k_1}} \sigma^{(k_2)}_{\alpha_{k_2}}+\cdots +
\\
\sum_{\{k_1,k_2,\cdots,k_M\}}\sum_{\alpha_{k_1}\alpha_{k_2}\cdots \alpha_{k_M}}t_{\alpha_{k_1}\alpha_{k_2}\cdots \alpha_{k_M}}\sigma^{(k_1)}_{\alpha_{k_1}} \sigma^{(k_2)}_{\alpha_{k_2}}\cdots \sigma^{(k_M)}_{\alpha_{k_M}}+ \cdots+
\\
\sum_{\alpha_{1}\alpha_{2}\cdots \alpha_{N}}t_{\alpha_{1}\alpha_{2}\cdots \alpha_{N}}\sigma^{(1)}_{\alpha_{1}} \sigma^{(2)}_{\alpha_{2}}\cdots \sigma^{(N)}_{\alpha_{N}}\} ,
\end{multline}
 where $\textbf{s}^{(k_1)}$ is a Bloch (coherent) vector corresponding to $k_1$th qubit, $\mf{s}^{(k_1)} =[s_{\alpha_{k_1}}]_{\alpha_{k_1}=1}^{3} ,$ which is a tensor of order one defined by
 \ben \l{e15}
 s_{\alpha_{k_1}}= tr[\rho \sigma^{(k_1)}_{\alpha_{k_1}}]= tr[\rho_{k_1} \sigma_{\alpha_{k_1}}],
  \een
 where $\rho_{k_1}$ is the reduced density matrix for the $k_1$th qubit. Here $\{k_1,k_2,\cdots,k_M\},\; 2 \le M \le N,$ is a subset of $\mathcal{N}$ and can be chosen in $\binom{N}{M}$  ways, contributing $\binom{N}{M}$ terms in the sum $\sum_{\{k_1,k_2,\cdots,k_M\}}$ in Eq.(\r{e14}), each containing a tensor of order $M$. The total number of terms in the Bloch representation of $\rho$ is $2^N$. We denote the tensors occurring in the sum $\sum_{\{k_1,k_2,\cdots,k_M\}},\; (2 \le M \le N)$ by $\mathcal{T}^{\{k_1,k_2,\cdots,k_M\}}=[t_{\alpha_{k_1}\alpha_{k_2}\cdots \alpha_{k_M}}]$ which  are defined by
\ben \l{e16}
 t_{\alpha_{k_1}\alpha_{k_2}\dots\alpha_{k_M}}= tr[\rho \sigma^{(k_1)}_{\alpha_{k_1}} \sigma^{(k_2)}_{\alpha_{k_2}}\cdots \sigma^{(k_M)}_{\alpha_{k_M}}]
 =tr[\rho_{k_1k_2\dots k_M} (\sigma_{\alpha_{k_1}}\otimes\sigma_{\alpha_{k_2}}\otimes\dots \otimes\sigma_{\alpha_{k_M}})]
\een
where $\rho_{k_1k_2\dots k_M}$ is the reduced density matrix for the subsystem $\{k_1 k_2\dots k_M\}$. We call the tensor in last term in Eq.(\r{e14}) $\mathcal{T}^{(N)}$.

In this article, we find the maximum in Eq.(\r{e9}) for a $N$-qubit state $\rho_{12\cdots N}$ to obtain an exact analytic formula, as in the two-qubit case (Eq.(\r{e3})) \cite{dakic}.

\emph{Theorem 2}. Let $\rho_{12\cdots N}$ be a $N$-qubit state defined by Eq.(\r{e14}), then
\ben \l{e17}
D_{k}(\rho_{12\cdots N})  = \frac{1}{2^N} \left[ ||\vec{s}^{(k)}||^2+\sum_{1 \leq M \leq N-1} \sum_{\{k_1,\ldots,k_M\} \in \mathcal{N}-k} ||\mathcal{T}^{\{k_1,\ldots,k_M,k\}}||^2-\eta_{max}\right].
\een
Here $\eta_{max}$ is the largest eigenvalue of the matrix $G^{(k)}$ which is a $3 \times 3$ real symmetric matrix, defined as
\ben \l{e18}
G^{(k)}=\vec{s}^{(k)}(\vec{s}^{(k)})^t+\sum_{k_1 \in \mathcal{N}-k}  (T^{\{k_1,k\}})^t T^{\{k_1,k\}}+\sum_{2\leq M \leq N-1} \mathbb{T}^{(M+1)} ,
\een
where $\mathbb{T}^{(M+1)}=[\tau^{M+1}_{\alpha_{k} \beta_{k}}]$ are $3 \times 3$ matrices, defined elementwise as
$$\tau^{(M+1)}_{\alpha_{k} \beta_{k}}=\sum_{\{k_1,k_2,\ldots,k_M\} \in \mathcal{N}-k}\sum_{\alpha_{k_1}\alpha_{k_2}\cdots \alpha_{k_M}} t_{\alpha_{k_1}\alpha_{k_2}\cdots \alpha_{k_M} \alpha_{k} } t_{\alpha_{k_1}\alpha_{k_2}\cdots \alpha_{k_M} \beta_{k}}$$
$$\alpha_{k_i},\al_{k} ,\beta_{k}=1,2,3;\;i=1,2,\ldots,M .$$

\emph{Proof}:

Our goal is to get a closed form expression for the term $\max_{A^{(k)}}||\mathcal{C}\times_{k} A^{(k)}||^2$ in Eq.(\r{e9}) applied to an arbitrary state $\rho_{12\cdots N}$ of a $N$-qubit system. The tensor $\mc{C}=[c_{i_1 i_2 \cdots i_N}]$ determined by the $N$-qubit state $\rho_{12\cdots N}$ via Eq.(\r{e4}) has $i_m = 1,2\; ; \; m=1,2,\ldots ,N,$ having $2^N$ elements in it. The $2\ti 4$ isometric matrices $A=[a_{li_k}]$ have to satisfy $a_{li_k}=tr(|l\rangle\langle l| X^{(k)}_{i_k}).$ In other words, the row vectors of $A^{(k)}$ must satisfy Eq.(\r{e12}) for some single qubit pure state. However, it is well known that every unit vector $\h{s}$ in ${\mb{R}}^3$ satisfies Eq.(\r{e12}) for some single qubit pure state (which is not true for a higher dimensional system \cite{ali1,kim}). Therefore, we can obtain the required maximum over {\it all} isometric $2\ti 4$ matrices in the form obtained below (See Eq.s(\r{eq711},\r{eq712})).      

We choose the orthonormal bases $\{X^{(m)}_{i_m}\},\; i_m=1,2,3,4;\;m=1,2,\ldots N$ in Eq.(\r{e4}) as the generators of $SU(2_m),\;m=1,2,\ldots,N$ [48].

\ben \l{eq112}
X^{(m)}_1=\frac{1}{\sqrt{2}}I_2,                   \\
\een
 and
 \ben \l{eq113}
 X^{(m)}_{i_m}= \frac{1}{\sqrt{2}}{\sigma}_{i_m-1},\; i_m=2,3,4;\;m=1,2,\ldots, N                \\
 \een
where ${\sigma}_{1,2,3}$ stand for the Pauli operators acting on the $m$th qubit.

 Since $tr{\sigma}_{\alpha_{k}}=0;\;\alpha_{k}=1,2,3$, we have,
 \ben
 \sum_{l=1}^{2} a_{li_{k}}=trX^{(k)}_{i_{k}}=\frac{1}{\sqrt{2}}tr{\sigma}_{i_{k}-1}=0,\; i_{k}=2,3,4. \nn \\
 \een
 Therefore,
 \ben  \l{e19}
 a_{2i_{k}}= - a_{1i_{k}},\;i_{k}=2,3,4.
\een

We now proceed to construct the $2\times 4$ matrix $A^{(k)}$ defined via Eq.(\r{e10}). We will use Eq.(\r{e19}). The row vectors of $A^{(k)}$ are
$$\vec{a}_l=(a_{l1},a_{l2},a_{l3},a_{l4}); l=1,2.$$
Next we define
\ben   \l{e20}
\h{e}_l=\sqrt{2}(a_{l2},a_{l3},,a_{l4}),\;l=1,2,
\een
and using Eq.(\r{e19}), we get
\ben   \l{e21}
\h{e}_{2}=- \h{e}_1.
\een

We can prove
\ben  \l{e22}
||\h{e}_l||^2=1 ,\;\;l=1,2 \; , 
\een
using the condition $||\vec{a}_l||^2=\sum_{i_{k}=1}^{4} a_{li_{k}}^2=1$ (Eq.(\r{eq901})) and using $a_{l1}= tr(|l\rangle \langle l| X^{(k)}_1)=\frac{1}{\sqrt{2}}.$

We can now construct the row vectors of $2\times 4$ matrix $A^{(k)}$, using Eq.(\r{e20}) and Eq.(\r{e21}).
\ben \l{eq711}
\vec{a}_1=\frac{1}{\sqrt{2}} (1, \h{e}_1),       \\
\een

\ben  \l{eq712}
\vec{a}_2=\frac{1}{\sqrt{2}} (1, -\h{e}_1) .             \\
\een

The matrix $A^{(k)}$ is, in terms of its row vectors defined above,

\begin{displaymath}
A^{(k)}=\frac{1}{\sqrt{2}}
\left(\begin{array}{cc}
1 & \h{e}_1\\
1 & -\h{e}_1\\
\end{array}\right).
\end{displaymath}






The norm of the tensor $\mc{C}$ can be expressed in terms of the norms of the tensors defining $\rho_{1\cdots N}$ by using the equivalence of the definitions of $\rho_{1\cdots N}$ given in Eq.(\r{e4}) and Eq.(\r{e14}) as 

\begin{widetext}
\begin{IEEEeqnarray}{rCl} \l{e23}
||\mathcal{C}||^2 & = & \frac{1}{2^N}\left[1+ \sum_{k_1 \in \mathcal{N}} ||\vec{s}^{(k_1)}||^2+\sum_{\{k_1,k_2\}} ||T^{\{k_1,k_2\}}||^2+\cdots \right.\nn \\
&& \:\left. +\sum_{\{k_1,k_2,\cdots,k_M\}}  ||T^{\{k_1,k_2,\cdots,k_M\}}||^2+\cdots+ + ||T^{(N)}||^2\right].
\end{IEEEeqnarray}
\end{widetext}

In order to get the norm of $\mathcal{C}\times_{k} A^{(k)}$ we use its elementwise definition,    
\begin{widetext}
\begin{IEEEeqnarray}{rCl}  \l{e24}
  (\mathcal{C}\times_{k} A^{(k)})_{i_1i_2\cdots i_{k-1} l i_{k+1}\cdots i_N} & = & \sum_{i_{k}} c_{i_1i_2\cdots i_{k-1} i_{k}i_{k+1}\cdots i_N} a_{li_{k}}, \nn \\
 && \: l=1,2
 \end{IEEEeqnarray}
\end{widetext}
the equivalence of the definitions of $\rho_{1\cdots N}$ given in Eq.(\r{e4}) and Eq.(\r{e14}) and the elements of $A^{(k)}$ given by Eq.s(\r{eq711},\r{eq712}). The result is

\begin{widetext}
\begin{IEEEeqnarray}{rCl}  \l{e29}
||\mathcal{C}\times_{k} A^{(k)}||^2 & = & \frac{1}{2^N}\left\{1+\sum_{k_1 \in\mathcal{N}-k} ||\vec{s}^{(k_1)}||^2 + \sum_{2\leq M \leq N-1} \sum_{\{k_1,k_2,\cdots,k_M\} \in \mathcal{N}-k}||T^{\{k_1,k_2,\cdots,k_M\}}||^2 \right. \nn \\
 && \: \left.+ \h{e}_1 \vec{s}^{(k)}(\vec{s}^{(k)})^t \h{e}_1^t +\sum_{k_1 \in\mathcal{N}-k}\h{e}_1 (T^{\{k_1,k\}})^t T^{\{k_1,k\}} \h{e}_1^t \right. \nn \\
  && \: \left. + \sum_{2\leq M \leq N-1} \sum_{\{k_1,\cdots,k_M\} \in \mathcal{N}-k}\sum_{\alpha_{k}=1}^3\sum_{\beta_{k}=1}^3 \h{e}_{1\alpha_{k}}(\sum_{\alpha_{k_1}\cdots \alpha_{k_M}} t_{\alpha_{k_1}\cdots \alpha_{k_M}\alpha_{k}} t_{\alpha_{k_1}\cdots \alpha_{k_M}\beta_{k}}) \h{e}_{1\beta_{k}} \right\} ,
 \end{IEEEeqnarray}
\end{widetext}

or,

\begin{widetext}
\begin{IEEEeqnarray}{rCl}  \l{e30}
||\mathcal{C}\times_{k} A^{(k)}||^2
 && \:= \frac{1}{2^N}\left\{ 1 +\sum_{k_1 \in\mathcal{N}-k} ||\vec{s}^{(k_1)}||^2 + \sum_{2\leq M \leq N-1} \sum_{\{k_1,k_2,\cdots,k_M\} \in \mathcal{N}-k}||T^{\{k_1,k_2,\cdots,k_M\}}||^2  \right.  \nn \\
 && \: \left. + \h{e}_1 \left[ \vec{s}^{(k)}(\vec{s}^{(k)})^t +\sum_{k_1 \in\mathcal{N}-k}(T^{\{k_1,k\}})^t T^{\{k_1,k\}} \right.\right. \nn \\
 && \:\left.\left. + \sum_{2\leq M \leq N-1} \mathbb{T}^{(M+1)} \right]\h{e}_1^t \right\} , 
\end{IEEEeqnarray}
\end{widetext}
where $\mathbb{T}^{(M+1)}=[\tau^{(M+1)}_{\alpha_{k} \beta_{k}}]$ with  $$\tau^{(M+1)}_{\alpha_{k} \beta_{k}}=\sum_{\{k_1,\cdots,k_M\} \in \mathcal{N}-k}\sum_{\alpha_{k_1}\alpha_{k_2}\cdots \alpha_{k_M}} t_{\alpha_{k_1}\alpha_{k_2}\cdots \alpha_{k_M} \alpha_{k} } t_{\alpha_{k_1}\alpha_{k_2}\cdots \alpha_{k_M} \beta_{k}} ,$$ as in the statement of the theorem.

Let us put the expression in square bracket, the  $(3\ti 3)$ real symmetric  matrix in Eq.(\r{e30}) as
\ben
G^{(k)}= \vec{s}^{(k)}(\vec{s}^{(k)})^t +\sum_{k_1 \in\mathcal{N}-k}(T^{\{k_1,k\}})^tT^{\{k_1,k\}}
 + \sum_{2\leq M \leq N-1} \mathbb{T}^{(M+1)}.  \nn
\een
Thus, we get
\begin{widetext}
\begin{IEEEeqnarray}{rCl}  \l{e31}
||\mathcal{C}\times_{k} A^{(k)}||^2
 && \:= \frac{1}{2^N}\left\{1 +\sum_{k_1 \in\mathcal{N}-k}||\vec{s}^{(k_1)}||^2  \right. \nn \\
 && \:\left. + \sum_{2\leq M \leq N-1} \sum_{\{k_1,k_2,\cdots,k_M\} \in \mathcal{N}-k}||T^{\{k_1,k_2,\cdots,k_M\}}||^2 + \h{e}_1 G^{(k)} \h{e}_1^t\right\}.
\end{IEEEeqnarray}
\end{widetext}

In Eq.(\r{e31}) only the last term depends on matrix $A^{(k)}$ while all others are determined by the state $\rho_{1\cdots N}.$ Therefore, to maximize $||\mathcal{C}\times_{k} A^{(k)}||^2$ we take $\h{e}_1$ to be the eigenvector of $G^{(k)}$ corresponding to its largest eigenvalue $\eta_{max},$ so that

\begin{widetext}
\begin{IEEEeqnarray}{rCl}\l{e32}
\max_{A^{(k)}}||\mathcal{C}\times_{k} A^{(k)}||^2
 && \: = \frac{1}{2^N}\left\{1 +\sum_{k_1 \in\mathcal{N}-k}||\vec{s}^{(k_1)}||^2 \nn  \right.  \\
 && \: \left. +\sum_{2\leq M \leq N-1} \sum_{\{k_1,k_2,\cdots,k_M\} \in \mathcal{N}-k}||T^{\{k_1,k_2,\cdots,k_M\}}||^2 + \eta_{max}\right\}.
 \end{IEEEeqnarray}
\end{widetext}

Finally, Eq.(\r{e23}), Eq.(\r{e32}) and  Eq.(\r{e9}) together imply
\begin{widetext}
\begin{IEEEeqnarray}{rCl}
D_{k}(\rho_{12\cdots N})=\frac{1}{2^N} \left\{||\vec{s}^{(k)}||^2+\sum_{1 \leq M \leq N-1} \sum_{\{k_1,k_2,\cdots,k_M\} \in \mathcal{N}-k} ||\mathcal{T}^{\{k_1,k_2,\cdots,k_M,k\}}||^2-\eta_{max}\right\}, \nn
 \end{IEEEeqnarray}
\end{widetext}
where $\eta_{max}$ is the largest eigenvalue of matrix $G^{(k)},$ thus completing the proof.

From Eq.(\r{e31}) and Eq.(\r{e32}) we note that the isometric $2\ti 4$ matrix ${\tilde{A}}^{(k)}$ which maximizes $||\mathcal{C}\times_{k} A^{(k)}||^2$ can now be explicitly constructed as
\begin{displaymath}
{\widetilde{A}}^{(k)}=\frac{1}{\sqrt{2}}
\left(\begin{array}{cc}
1 & \h{e}_{max}\\
1 & -\h{e}_{max}\\
\end{array}\right),
\end{displaymath}
where ${\h{e}}_{max}$ is the eigenvector of $G^{(k)}$ for its highest eigenvalue $\eta_{max}.$ We can then use Eq.(\r{e9}) directly to compute 
\ben \l{eq1}
D_{k}(\rho_{12\cdots N}) = ||\mathcal{C}||^2- ||\mathcal{C}\times_{k} {\widetilde{A}}^{(k)}||^2 .     \\
\een

For a two qubit system Eq.(\r{e17}) reduces to 
\ben \l{eq4}
D_{1}(\rho_{12})=\frac{1}{4} (||\vec{x}||^2+ ||\mathcal{T}||^2-\eta_{max}),   \\
\een
and
\ben \l{eq5}
D_{2}(\rho_{12})=\frac{1}{4} (||\vec{y}||^2+ ||\mathcal{T}||^2-\zeta_{max}),    \\
\een
where $\vec{x}, \vec{y}$ are the coherent vectors of the reduced density operators of the first and the second qubit respectively, $\mc{T}$ is the two qubit correlation matrix and $\eta_{max},\zeta_{max}$ are the largest eigenvalues of $G^{(1)}=\v{x}\v{x}^t+\mc{T}\mc{T}^t$ and $G^{(2)}=\v{y}\v{y}^t+\mc{T}^t\mc{T}$ respectively.   

Interestingly, the quantum discord $D_{k}(\rho_{12\cdots N})$ can be obtained experimentally, without a detailed knowledge of the state $\rho_{12\cdots N},$
because all the elements of the matrix $G^{(k)}$ as well as the tensor $\mc{C}$ (Eq.(\r{e4})) (both of which are the average values of the tensor products of Pauli operators in the $N$-qubit state) can be experimentally determined by measuring Pauli operators on individual qubits.    \\  

\textbf{IV. EXAMPLES}\\

Next, we apply our measure to some multiqubit quantum states. Unfortunately, a quantitative comparison with the entanglement-separability scenario still eludes us because a viable measure of entanglement for multipartite mixed states is not available. 

The first example comprises the 3-qubit mixed states 
\ben \l{eqn1}
\rho = p |GHZ\rangle\langle GHZ|+\frac{(1-p)}{8} I_8,\; 0\leq p \leq 1;
\een
where $|GHZ \rangle =\frac{1}{\sqrt{2}}(|000\rangle+|111\rangle)$ and $I_8$ is the identity matrix. Figure 1(a) shows the variation of $D_1(\rho)$ with $p$. We see that $D_1(\rho)$ increases continuously from $p=0$ state (random mixture) to $p=1$ state (pure GHZ state), as expected.  

\begin{figure}[!ht]
\begin{center}

\includegraphics[width=8cm,height=6cm]{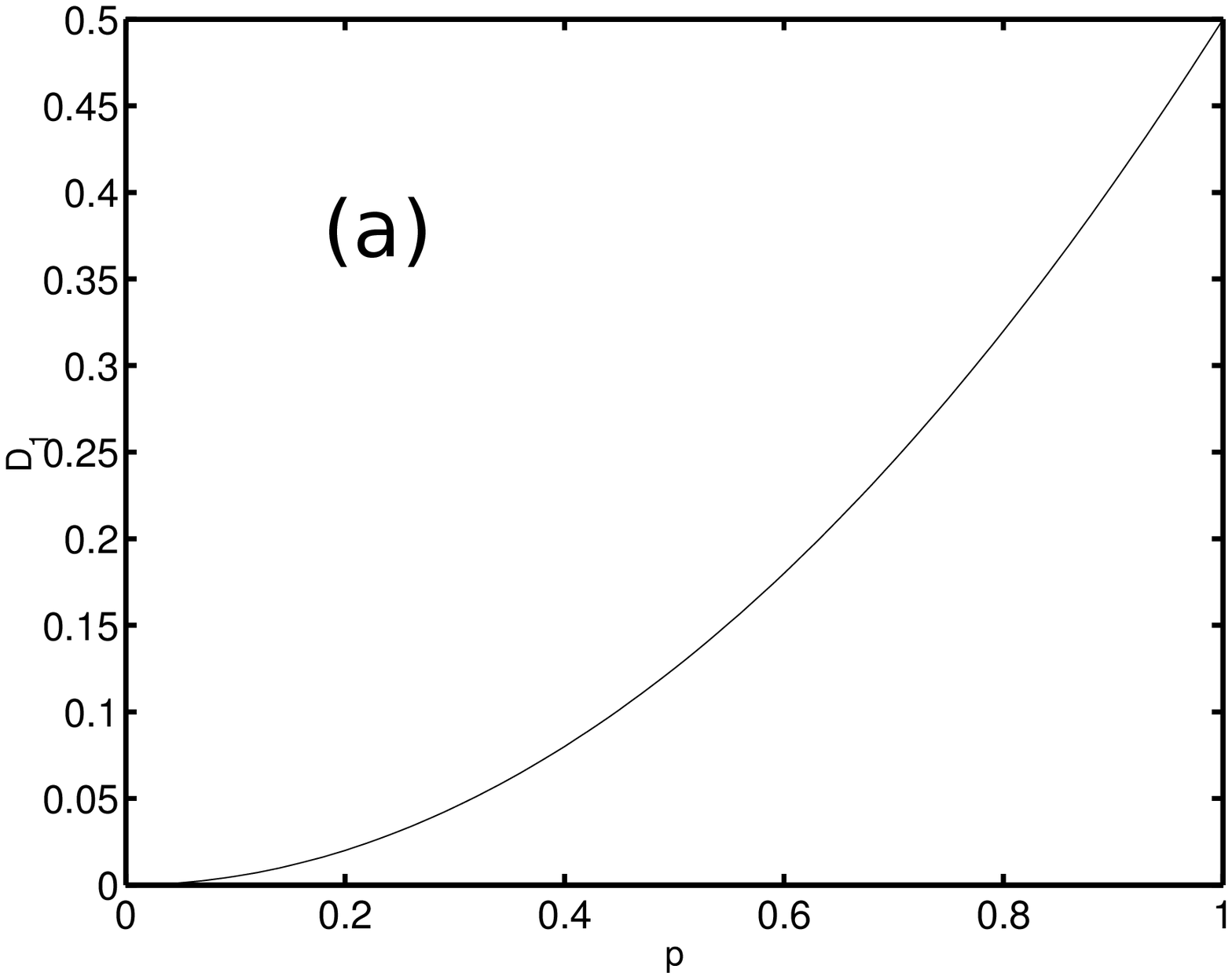}

 
\end{center}
\end{figure}

\begin{figure}[!ht]
\begin{center}

\includegraphics[width=8cm,height=6cm]{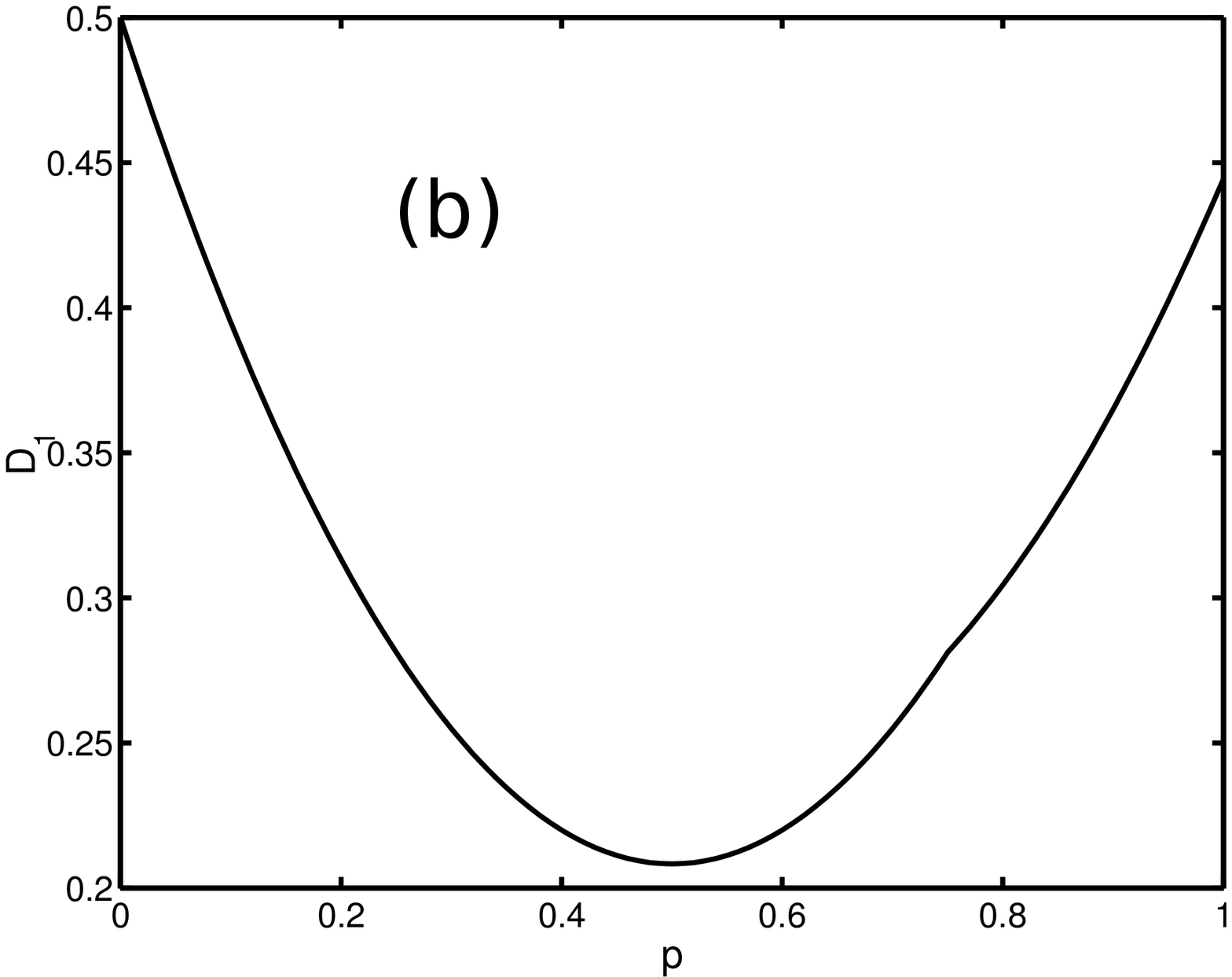}


\end{center}
\end{figure}

\begin{figure}[!ht]
\begin{center}

\includegraphics[width=8cm,height=6cm]{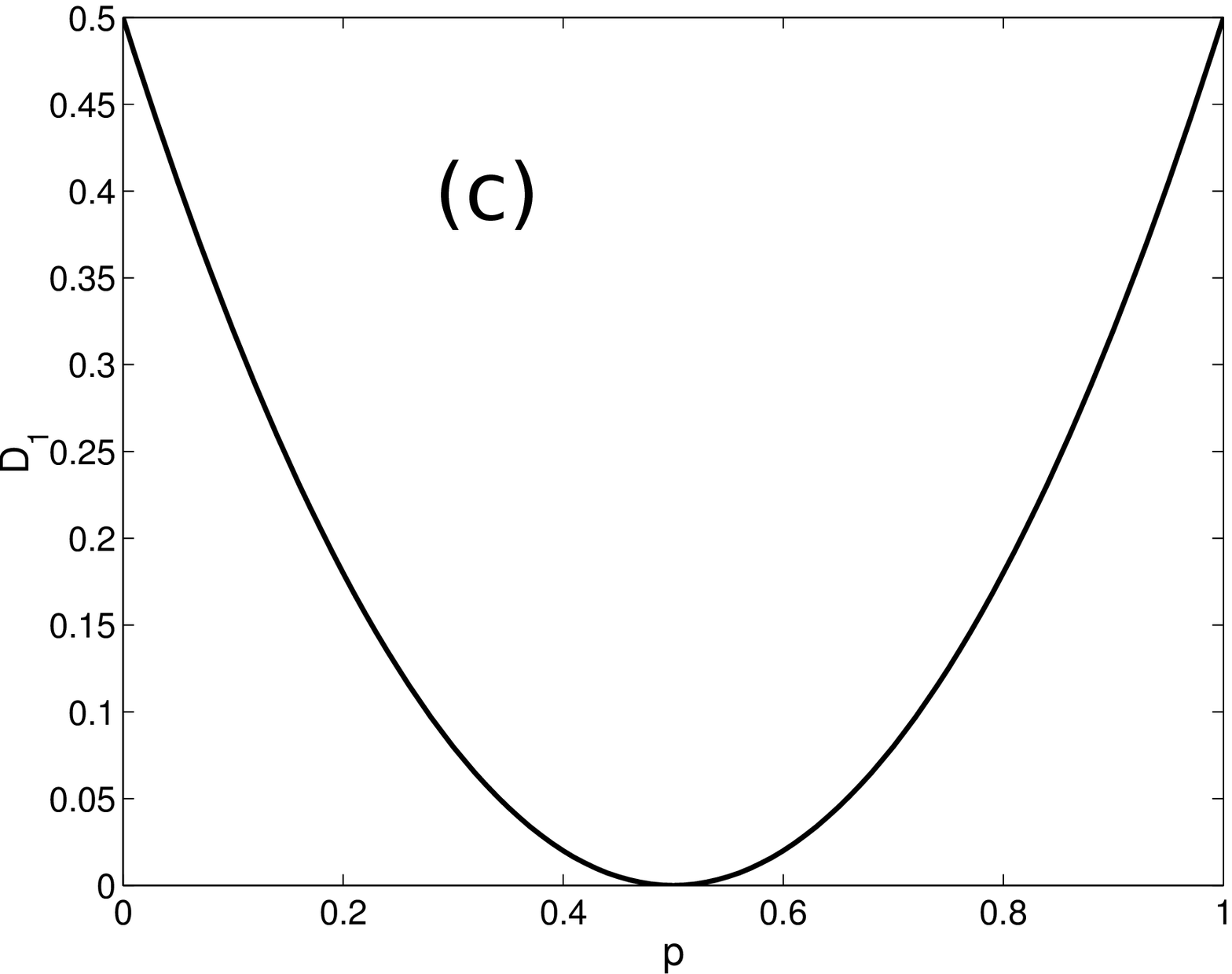}

Figure 1 : Variation of the quantum discord with parameter $p$ for the states given in (a) Eq.(\r{eqn1}) (b) Eq.(\r{eqn2}) and (c)Eq.(\r{eqn3}).
\end{center}
\end{figure}

Second example is the set of 3-qubit states 
\ben \l{eqn2}
\rho= p |W\rangle\langle W|+(1-p) |GHZ\rangle\langle GHZ|,\; 0\leq p \leq 1;
\een
where $|W\rangle=\frac{1}{\sqrt{3}}(|100\rangle+|010\rangle +|001\rangle)$. Figure 1(b) shows the variation of $D_1(\rho)$ with $p$. It is straightforward to check that this state cannot be written as a classical quantum state for any value of $p,$ including $p=\frac{1}{2}.$ This explains the nonzero discord at $p=\frac{1}{2}.$ Further, we observe that discord for the pure GHZ state exceeds that for the pure W state, in conformity with similar behavior of entanglement in these states \cite{joag}. The rate of increase of the discord diminishes discontinuously at $p=\frac{3}{4}$ as the $|W\ran$ state increasingly dominates the classical mixture with increasing $p.$ This interesting observation needs further analysis.    

As the last example we consider the set of 3-qubit states 
\ben \l{eqn3}
\rho= p |GHZ_-\rangle\langle GHZ_-|+(1-p) |GHZ\rangle\langle GHZ|,\; 0\leq p \leq 1;
\een
where $|GHZ_-\rangle=\frac{1}{\sqrt{2}}(|000\rangle-|111\rangle)$. Figure 1(c) shows the variation of $D_1(\rho)$ with $p$. The discord is symmetric about $p=\frac{1}{2}$ at which it vanishes. For $p=\frac{1}{2}$ the state can be written as $$\frac{1}{2}|000\ran\lan 000|+\frac{1}{2}|111\ran\lan 111|$$ which is a classical quantum state, so that discord vanishes at $p=\frac{1}{2}.$ Again, discord is maximum and equal for pure $|GHZ\ran$ state and pure $|GHZ_{-}\ran$ state, similar to the behavior of entanglement in these two states \cite{joag}.

We note that, in all these examples, $D_1(\rho)=D_2(\rho)=D_3(\rho)$ as all the states are symmetric with respect to the swapping of qubits.     \\

\textbf{V. TOTAL QUANTUM CORRELATIONS IN A BIPARTITE STATE}\\

Consider a bipartite state $\rho$ and denote by ${\widetilde{\Pi}}^{(1)}$ the von Neumann measurement minimizing $||\rho-\Pi^{(1)}(\rho)||^2.$
It is straightforward to check that the state after the measurement ${\widetilde{\Pi}}^{(1)}(\rho)$ is a zero discord state, that is $D_1( {\widetilde{\Pi}}^{(1)}(\rho))=0.$
However, the state $\widetilde{\Pi}^{(1)}(\rho)$ may have $D_2( {\widetilde{\Pi}}^{(1)}(\rho))\ne 0.$ Thus the state $\widetilde{\Pi}^{(1)}(\rho)$ can have some non-zero quantum correlations. Thus neither $D_1(\rho)$ nor $D_2(\rho)$ gives us a measure of the total quantum correlations in the state $\rho$. But this analysis suggests that the quantity

\ben \l{eq6}
Q(\rho)= D_1(\rho)+D_2(\widetilde{\Pi}^{(1)}(\rho))    \\
\een
 gives the required measure of the total quantum correlations in the state $\rho$ \cite{okrasa}.

 In order to find the optimal von Neumann measurement $\widetilde{\Pi}^{(1)}$ on $\rho$ which minimizes $||\rho-\widetilde{\Pi}^{(1)}(\rho)||^2$ we have to find the corresponding orthonormal basis $\{|\widetilde{q}\rangle\}$ in $\mc{H}^1$ such that $\{\widetilde{\Pi}^{(1)}_q\}=\{|\widetilde{q}\rangle\langle \widetilde{q}|\}$. The expansion of these 1-D projectors $|\widetilde{q}\rangle\langle \widetilde{q}|$ in the basis $X_i=\{I_1,\lambda_i\}$ ($\lambda_i$ : generators of $SU(d_1)$) via Eq.(\r{eq801}), that is,
 \ben \l{eq25}
 |\widetilde{q}\ran\lan \widetilde{q}|=\sum_i^{d_1^2} {\widetilde{a}}_{qi}X_i\;\;; q=1,\ldots,d_1 \\
 \een
 with
 \ben \l{eq26}
 {\widetilde{a}}_{qi}=\lan \widetilde{q}|X_i|\widetilde{q}\ran,\;\;  q = 1,2\ldots,d_1;\;i = 1,\ldots,d_1^2 ,     \\
 \een
 must then give the matrix ${\widetilde{A}}^{(1)}$ which maximizes $tr(ACC^tA^t)$ which in turn gives $D_1(\rho).$

To get the state $\widetilde{\Pi}^{(1)}(\rho)$ we proceed as follows. As noticed above, any post measurement state ${\Pi}^{(1)}(\rho)$ is a zero discord state satisfying $D_1({\Pi}^{(1)}(\rho))=0$. Hence ${\Pi}^{(1)}(\rho)$ must have the form of classical quantum state as in Eq.(\r{e6}) for $N=2$ namely,
\ben \l{eq111}
{\Pi}^{(1)}(\rho)=\sum_{q=1}^{d_1}p_q|q\ran\lan q|\ot \rho_q  .   \\
\een
We expand the state $p_q \rho_q$ in Eq.(\r{eq111}) in terms of the basis $\{X^{(2)}_j\}$ to get
\ben \l{eq27}
p_q \rho_q=\sum_j b_{qj} X^{(2)}_j ,   \\
\een
 where $b_{qj}=tr(p_q \rho_q X^{(2)}_j)$. We know from ref. \cite{fu} that, for Eq.(\r{eq1}) for $N=2$ to hold, we must have
\ben \l{eq28}
b_{qj}=\sum_i{\widetilde{a}}_{qi}c_{ij}.
\een
where the matrix $C=[c_{ij}]$ is defined via Eq.(\r{e4}) for $N=2.$ Now, we substitute Eq.(\r{eq25}), Eq.(\r{eq27}) and Eq.(\r{eq28}) in the expression for the general post measurement state (Eq.(\r{eq111})) to get the state $\widetilde{\Pi}^{(1)}(\rho)$ which easily reduces to
\ben \l{eq29}
\widetilde{\Pi}^{(1)}(\rho)=\sum_{lj} (\widetilde{A}^{(1)t} \widetilde{A}^{(1)} C)_{lj} X^{(1)}_l \otimes X^{(2)}_j ,
\een
where $\widetilde{A}^{(1)}$ is the matrix which maximizes $tr(A^{(1)}CC^tA^{(1)t}).$ Thus $\widetilde{\Pi}^{(1)}(\rho)$ has the form
$$\widetilde{\Pi}^{(1)}(\rho)=\sum_{lj} {c^{\pr}}_{lj} X^{(1)}_l \otimes X^{(2)}_j$$  same as in Eq.(\r{e4}) for $N=2$. 

 Specializing to $2$-qubit systems, we have, for $D_2(\widetilde{\Pi}^{(1)}(\rho)),$ using Eq.(\r{eq5}),
\ben \l{eq30}
D_2(\widetilde{\Pi}^{(1)}(\rho))= \frac{1}{4}(||\vec{y}||^2+||\widetilde{T}||^2-\widetilde{\zeta}_{max}) ,
\een
where $y_j=tr(I\otimes \sigma_j \widetilde{\Pi}^{(1)}(\rho))= tr(I\otimes \sigma_j \rho)$, $\widetilde{t}_{ij}=tr (\sigma_i\otimes \sigma_j \widetilde{\Pi}^{(1)}(\rho)),$ ${\widetilde{\zeta}}_{max}$ is the largest eigenvalue of the real symmetric matrix $$ {\widetilde{G}}^{(2)}=\vec{y}\vec{y}^t+ \widetilde{T}^t \widetilde{T},$$
and $\widetilde{\Pi}^{(1)}(\rho)$ is given by Eq.(\r{eq29}).
The total quantum correlations in the state $\rho$ are given by
\ben \l{eq31}
Q(\rho)= D_1(\rho)+D_2(\widetilde{\Pi}^{(1)}(\rho)))
\een
along with equations Eq.(\r{eq4}), Eq.(\r{eq30}) and Eq.(\r{eq29}) (for $\widetilde{\Pi}^{(1)}(\rho)$).\\

\textbf{VI. TOTAL QUANTUM CORRELATIONS IN A $N$-PARTITE STATE}\\

In this section, we obtain a closed form expression for the total quantum correlations in a $N$-partite quantum state. We use Eq. (\r{e9},\r{e11}).

Consider a N-partite state $\rho_{12\cdots N}$ and denote by ${\widetilde{\Pi}}^{(k)}$ the von Neumann measurement minimizing Eq.(\r{e11}).
It is straightforward to check that the state after the measurement ${\widetilde{\Pi}}^{(k)}(\rho_{12\cdots N})$ is a zero $k$-discord state, that is $D_k( {\widetilde{\Pi}}^{(k)}(\rho_{12\cdots N}))=0.$
However, the state ${\widetilde{\Pi}}^{(k)}(\rho_{12\cdots N})$ may have $D_l( {\widetilde{\Pi}}^{(k)}(\rho_{12\cdots N}))\ne 0,\; l \neq k.$ Thus the state ${\widetilde{\Pi}}^{(k)}(\rho_{12\cdots N})$ can have some non-zero quantum correlations. Thus $D_k(\rho_{12\cdots N})$ cannot give us a measure of the total quantum correlations in the state $\rho_{12\cdots N}.$ This analysis suggests a geometric measure of total quantum correlations present in a $N$-partite state $\rho_{12\cdots N}$ \cite{okrasa}.

We can now use the above considerations to investigate the total quantum correlations present in a state $\rho_{12\cdots N}$.
Let us assume that the non-selective von Neumann projective measurements ${\widetilde{\Pi}}^{(1)},{\widetilde{\Pi}}^{(2)},\cdots,{\widetilde{\Pi}}^{(N)}$ are performed successively on $N$ parts $12\cdots N ,$ $k$th successive measurement being performed on the $k$th part, leading to $D_k( \mu_{12\cdots N})=0 ,$ where $\mu_{12\cdots N}$ is the state produced after $(k-1)$th successive measurement, given in Eq.(\r{eq33}). Clearly, the corresponding post-measurement states are given by
\ben \l{eq33}
{\widetilde{\Pi}}^{(1)}(\rho_{12\cdots N}), {\widetilde{\Pi}}^{(2)}({\widetilde{\Pi}}^{(1)}(\rho_{12\cdots N})), \ldots, {\widetilde{\Pi}}^{(N)}(\cdots({\widetilde{\Pi}}^{(1)}(\rho_{12\cdots N})\cdots) .
\een
Here the measurement ${\widetilde{\Pi}}^{(k)}$ minimizes the loss of correlations in the state produced after the first $k-1$ successive measurements on $k-1$ parts. Thus the geometric measures of quantum discord of these successive measurement states are given by
\ben \l{eq34}
D_1(\rho_{12\cdots N}),
\een
\ben \l{eq35}
D_2( {\widetilde{\Pi}}^{(1)}(\rho_{12\cdots N})),
\een
\ben \l{eq36}
D_3( {\widetilde{\Pi}}^{(2)}({\widetilde{\Pi}}^{(1)}(\rho_{12\cdots N}))),
\een
$$\vdots$$
\ben \l{eq37}
D_N( {\widetilde{\Pi}}^{(N-1)}(\cdots({\widetilde{\Pi}}^{(1)}(\rho_{12\cdots N})))\cdots).
\een
Therefore, the geometric measure of total quantum correlations present in a
N-partite quantum state $\rho_{12\cdots N}$ is given by

\begin{multline} 
Q(\rho_{12\cdots N})= D_1(\rho_{12\cdots N})+D_2( {\widetilde{\Pi}}^{(1)}(\rho_{12\cdots N}))+D_3( {\widetilde{\Pi}}^{(2)}({\widetilde{\Pi}}^{(1)}(\rho_{12\cdots N})))+\cdots
   \\
\cdots +D_N( {\widetilde{\Pi}}^{(N-1)}(\cdots({\widetilde{\Pi}}^{(1)}(\rho_{12\cdots N})))\cdots),
\end{multline}
which is a multipartite generalization of the measure (\r{eq31}) introduced in the previous section.

We use Eq.(\r{e9}) to write, for the quantum discord $D_k$ corresponding to the $k$th successive measurement on the $k$th part,  
\ben \l{eq200}
D_k( {\widetilde{\Pi}}^{(k-1)}(\cdots({\widetilde{\Pi}}^{(1)}(\rho_{12\cdots N})))\cdots)= ||\mc{C}^{(k)}||^2 - ||\mc{C}^{(k)} \ti_k \widetilde{A}^{(k)}||^2 , 
\een
where $\widetilde{A}^{(k)}$ gives the maximum value of the second term. From Eq.(A4) in the Appendix we can infer that
\ben \l{eq201}
\mc{C}^{(k)}= \mc{C}^{(k-1)}\ti_{k-1} (\widetilde{A}^{(k-1)t} \widetilde{A}^{(k-1)})\;\;\; k=2,3,\ldots,N .    \\
\een
 Therefore, we get, for the total quantum correlations $Q(\rho_{12\cdots N}),$ 
\benr  \l{eq219}
Q(\rho_{12\cdots N})&=&||\mathcal{C}^{(1)}||^2 - ||\mathcal{C}^{(1)}\times_{1} {\widetilde{A}}^{(1)}||^2 \nn   \\
&+& \sum_{k=2}^{N}(||\mathcal{C}^{(k-1)}\ti_{k-1} (\widetilde{A}^{(k-1)t} \widetilde{A}^{(k-1)})||^2 - ||\mathcal{C}^{(k-1)}\ti_{k-1} (\widetilde{A}^{(k-1)t} \widetilde{A}^{(k-1)})\ti_k \widetilde{A}^{(k)}||^2) .  \nn  \\
\eenr
Written explicitly, the $k$th term in the expression of $Q(\rho_{12\cdots N})$ for $2\le k\le N$ is,
$$||\mathcal{C}^{(1)}\Pi_{j=2}^{k}\times_{j-1}((\widetilde{A}^{(j-1)t}) \widetilde{A}^{(j-1)}||^2 - ||\mathcal{C}^{(1)}(\Pi_{j=2}^{k}\times_{j-1}((\widetilde{A}^{(j-1)t}) \widetilde{A}^{(j-1)})\ti_k \widetilde{A}^{(k)}||^2,$$
where $\mathcal{C}^{(1)}=\mc{C}$ and $\widetilde{A}^{(1)},$ maximizing the second term in Eq.(\r{e9}), correspond to the starting state $\rho_{12\cdots N}$ via Eq.s(\r{e4},\r{e9}). 


Next we prove that $$||\mathcal{C}^{(k-1)}\times_{k-1} \widetilde{A}^{(k-1)}||^2 = ||\mc{C}^{(k)}||^2 = ||\mathcal{C}^{(k-1)}\times_{k-1} \widetilde{A}^{(k-1)t}\widetilde{A}^{(k-1)}||^2,\;k=2,\ldots,N.$$
 Using the definition of norm of a tensor as the inner product of a tensor with itself given in \cite{kold06} we get,  

\ben
 ||\mathcal{C}^{(k-1)}\times_{k-1}((\widetilde{A}^{(k-1)t}) \widetilde{A}^{(k-1)})||^2 = \langle \mathcal{C}^{(k-1)}\times_{k-1}((\widetilde{A}^{(k-1)t})\widetilde{A}^{(k-1)}) ,  \mathcal{C}^{(k-1)}\times_{k-1}((\widetilde{A}^{(k-1)t})\widetilde{A}^{(k-1)})\rangle, \nn
 \een
 We use Proposition 3.11  and  Proposition 3.4(b) in \cite{kold06} to get
 \begin{multline}
\langle \mathcal{C}^{(k-1)}\times_{k-1}((\widetilde{A}^{(k-1)t})\widetilde{A}^{(k-1)}) ,  \mathcal{C}^{(k-1)}\times_{k-1}((\widetilde{A}^{(k-1)t})\widetilde{A}^{(k-1)})\rangle   
   \\
 = \langle \mathcal{C}^{(k-1)}\times_{k-1}((\widetilde{A}^{(k-1)t}) \widetilde{A}^{(k-1)}) \times_{k-1}((\widetilde{A}^{(k-1)t}) \widetilde{A}^{(k-1)}), \mathcal{C}^{(k-1)} \rangle  
   \\     
= \langle \mathcal{C}^{(k-1)}\times_{k-1}((\widetilde{A}^{(k-1)t})(\widetilde{A}^{(k-1)}\widetilde{A}^{(k-1)t}) \widetilde{A}^{(k-1)}), \mathcal{C} \rangle \nn  \\
\end{multline}
 We know that $\widetilde{A}^{(k)}(\widetilde{A}^{(k)})^t=I \;; k=1,\ldots,N$, so that,
 \benr
||\mathcal{C}^{(k-1)}\times_{k-1}((\widetilde{A}^{(k-1)t}) \widetilde{A}^{(k-1)})||^2 &=& \langle \mathcal{C}^{(k-1)}\times_{k-1}((\widetilde{A}^{(k-1)t}) \widetilde{A}^{(k-1)}), \mathcal{C}^{(k-1)} \rangle \nn   \\
&=&  \langle \mathcal{C}^{(k-1)}\times_{k-1} \widetilde{A}^{(k-1)}, \mathcal{C}^{(k-1)}\times \widetilde{A}^{(k-1)}  \rangle =||\mathcal{C}^{(k-1)}\times_{k-1} \widetilde{A}^{(k-1)}||^2 . \nn  \\
 \eenr

By using this, all the terms except the first and the last term in Eq.(\r{eq219}) pairwise cancel . Thus, we finally get
\ben \l{eq40}
Q(\rho_{12\cdots N})=||\mathcal{C}||^2 - ||\mathcal{C}\times_1 \widetilde{A}^{(1)}\times_2 \widetilde{A}^{(2)}\times_3 \cdots \times_{N-1}  \widetilde{A}^{(N-1)}\times_{N} {\widetilde{A}}^{(N)}||^2 . \\
\een
 This formula applies to an arbitrary $N$-partite quantum state. However, $Q(\rho_{12\cdots N})$ can be actually computed only for a $N$-qubit state, because the matrices $\widetilde{A}^{(k)}$ as well as the states $\widetilde{\Pi}^{(k)}(\rho_{12\cdots N}), k=1,\ldots, N$ can be explicitly constructed in this case, as shown in section III and the appendix (Eq.(A4)). Further, for $N$-qubit states, this formula can be experimentally implemented, as all the elements of all the matrices can be determined by measuring Pauli operators on individual qubits.
 
 From Proposition 3.4(a) of ref.\cite{kold06}, namely,
\ben
\mathcal{Y}\times_m A\times_n B = (\mathcal{Y}\times_m A)\times_n B = (\mathcal{Y}\times_n B)\times_m A \;\;\;m\ne n ,   \nn   \\
\een
where $\mc{Y}\in\mb{R}^{J_1\times J_2\times \cdots \ti J_N}$ is a $N$-way tensor and $A\in \mb{R}^{L_{m}\ti J_{m}},$ $B\in \mb{R}^{L_{n}\ti J_{n}}$ are matrices, it follows that the total quantum correlation is invariant under arbitrary permutation of factors in Eq.(\r{eq40}) or is invariant under any permutation of order in which the measurements on individual parts are made.\\

\textbf{VII. SUMMARY AND COMMENTS}\\

To summarize, we obtain generic forms of quantum discord and total quantum correlations in a $N$-partite state and the corresponding exact formulas in the $N$-qubit case. The formulas for the quantum discord and the total quantum correlations in the $N$-qubit case are not only exactly computable using the $N$-qubit quantum state, but can also be experimentally implemented when the $N$-qubit state is not precisely known. States of quantum systems may not be known at some intermediate stage of quantum information processing and deciphering an unknown quantum state is a formidable task. Hence it is of great advantage if the crucial resourses like entanglement or quantum discord can be estimated experimentally, without taking recourse to what the quantum state is. The number of quantities required to be measured goes linearly with the system size $N,$ as only three Pauli operators are to be measured on a qubit. The computational complexity of the discord in a $N$-qubit state is dominated by that of the middle term in Eq.(\r{e17}) which deals with $3\ti 4^{N-1}$ elements (average values of the tensor products of Pauli operators in the $N$-qubit state). Thus computation increases exponentially with system size. This is not a real restriction when $N$ is small ($N=2,3,4$ qubits). Regarding the operational procedure defining the total quantum correlation in a $N$-partite state, we note that it accounts for the total quantum correlations between $N$ parts as no quantum correlations remain, including all possible cuts. However, if one or more parts contain correlated sub-parts and the optimal measurement is a joint measurement (in an entangled basis) on these sub-parts, then the correlations between these sub-parts remain. These correlations may be eliminated by considering the sub-parts as separate parts of the system. We do not lose any generality in this situation because we are concerned with the quantum correlations between the specified $N$ parts and our formulas account for these correlations. The actual division of the system into subsystems (parts) is dictated by the correlations required for a particular application. Finally, it will be interesting to seek a generalization of this work to include POVMs.        

\emph{Acknowledgments }:

This work was supported by the BCUD research grant RG-13. ASMH  thanks Pune University for the hospitality during his visit when this work was initiated. PSJ thanks Anil Shaji and Sai Vinjanampathy for a useful discussion. \\

\emph{Appendix : Finding the state} $\widetilde{\Pi}^{(k)}(\rho_{12\cdots N})$

 In order to find the optimal von Neumann measurement $\widetilde{\Pi}^{(k)}$ on $\rho_{12\cdots N}$ which minimizes $||\rho_{12\cdots N}-{\Pi}^{(k)}(\rho_{12\cdots N})||^2$ we have to find the corresponding orthonormal basis $\{|\widetilde{z}\rangle\}$ in $H^k$ such that $\{\widetilde{\Pi}^{(k)}_z\}=\{|\widetilde{z}\rangle\langle \widetilde{z}|\}$. The expansion of these 1-D projectors $|\widetilde{z}\rangle\langle \widetilde{z}|$ in the basis $X^{(k)}_i \;i=1,\ldots,d_{k}^2$ ($X^{(k)}_i$ : generators of $SU(d_k)$) that is,
 $$|\widetilde{z}\ran\lan \widetilde{z}|=\sum_i {\widetilde{a}}_{zi}X^{(k)}_i\;\;; i=1,\ldots,d_k^2 \eqno{(A1)}$$
  with
 $$ {\widetilde{a}}_{zi}=\lan \widetilde{z}|X^{(k)}_i|\widetilde{z}\ran,\;\;  z = 1,2,\ldots,d_k;\;i = 1,2,\ldots,d_k^2.$$
 must then give the matrix $\widetilde{A}^{(k)}$ which maximizes $||\mathcal{C}\times_k A^{(k)}||^2$ which in turn gives the $k$-discord $D_k(\rho_{12\cdots N}).$

To get the state $\widetilde{\Pi}^{(k)}(\rho_{12\cdots N})$ we proceed as follows. As noticed above, the state $\widetilde{\Pi}^{(k)}(\rho_{12\cdots N})$ is a zero $k$-discord state satisfying $D_k(\widetilde{\Pi}^{(k)}(\rho_{12\cdots N}))=0$. Hence $\widetilde{\Pi}^{(k)}(\rho_{12\cdots N})$ must have the form of classical quantum state as in Eq.(\r{e6}). We expand the state $p_z \rho_{[k]|z}$ in Eq. (\r{e6}) in terms of the basis $\{X^{(1)}_j\otimes X^{(2)}_j\otimes \cdots X^{(k-1)}_j \otimes X^{(k+1)}_j\otimes \cdots X^{(N)}_j\}$ to get,

 $$p_z\rho_{[k]|z}=\sum_{i_1i_2\cdots i_{k-1} i_{k+1} \cdots i_N} b_{i_1i_2\cdots i_{k-1} z i_{k+1} \cdots i_N} X^{(1)}_{i_1}\otimes X^{(2)}_{i_2}\otimes \cdots X_{i_{k-1}}^{(k-1)} \otimes X_{i_{k+1}}^{(k+1)}\otimes \cdots X_{i_N}^{(N)},$$
$$ z=1,2,\ldots,d_k;\; i_{m}=1,\ldots,d_k^2;\; m=1,2,\cdots, N, \eqno{(A2)}$$
 with $b_{i_1i_2\cdots i_{k-1} l i_{k+1} \cdots i_N}= tr(p_l\rho_{[k]|l} X^{(1)}_{i_1}\otimes X^{(2)}_{i_2}\otimes \cdots X_{i_{k-1}}^{(k-1)} \otimes X_{i_{k+1}}^{(k+1)}\otimes \cdots \ot X_{i_N}^{(N)}).$ We know from theorem 1 that, for Eq.(\r{e9}) to hold, we must have
$$b_{i_1i_2\cdots i_{k-1} z i_{k+1} \cdots i_N}=\sum_{i_{k}} c_{i_1i_2\cdots  i_N} \widetilde{a}_{zi_{k}}\eqno{(A3)}$$

Now, we substitute Eq.(A1), Eq.(A2) and Eq.(A3) in the expression for the general post measurement state (Eq.(\r{e6}))
$$\chi_{k}=\sum_{z=1}^{d_k} p_z |z\rangle \langle z|\otimes \rho_{[k]|z},$$
and use the definition of n-mode product in Eq.(\r{e7}) and  proposition 3.4(b) in ref.\cite{kold06} to get the state $\widetilde{\Pi}^{(k)}(\rho_{12\cdots N})$ which easily reduces to

$$\widetilde{\Pi}^{(k)}(\rho_{12\cdots N})= \sum_{i_1 i_2 \cdots i_N} [\mathcal{C}\times_k((\widetilde{A}^{(k)})^t \widetilde{A}^{(k)})]_{i_1 i_2 \cdots i_N} X^{(1)}_{i_1}\otimes X^{(2)}_{i_2} \otimes \cdots \otimes X^{(N)}_{i_N}  \eqno{(A4)}$$

where $\widetilde{A}^{(k)}$ is the matrix which maximizes $||\mathcal{C}\times_k A^{(k)}||^2.$ Thus $\widetilde{\Pi}^{(k)}(\rho_{12\cdots N})$ has the form
$$\widetilde{\Pi}^{(k)}(\rho_{12\cdots N})=\sum_{i_1 i_2 \cdots i_N} {\mathcal{C}^{\pr}}_{i_1 i_2 \cdots i_N} X^{(1)}_{i_1}\otimes X^{(2)}_{i_2} \otimes \cdots \otimes X^{(N)}_{i_N}$$

 same as in Eq.(\r{e4}).\\


\begin{thebibliography}{99}
\bibitem{com}
This amounts to choosing for $\sum_{l}p_{l}\rho_{[k]|l}$ the reduced density operator obtained by tracing out the $k$th part from the state $\rho_{12\cdots N}$ (Eq.(\r{e4})), 
which can always be done because $\{\rho_{[k]|l}\}$ is a set of arbitrary density operators acting on $\mathcal{H}^{1}\otimes \mathcal{H}^{2} \otimes \cdots \mathcal{H}^{k-1} \otimes\mathcal{H}^{k+1}\otimes \cdots \mathcal{H}^{N}$ and $\{p_l\}$ is an arbitrary probability distribution.
\end{thebibliography}
\end{document}